\documentclass[10pt,draft]{article}
\usepackage{epsf,amsmath}

\begin{document}

\hfill LA-UR-03-5879

\begin{center}
{\large \bf Dilepton Transverse Momentum in the Color Dipole Approach
\footnote{Talk presented by M. B. Gay Ducati at  DIS03, partially
  supported by PRONEX-CNPq. M. A. Betemps and M. V.  Machado were 
  supported by CNPq (Brazil). J. Raufeisen was supported by the US
  Department of Energy at Los Alamos under Contract No.\
  W-7405-ENG-38.}}

M. A. Betemps$^a$, M. B. Gay Ducati$^a$, M. V. T. Machado$^{a,b}$,\\  
J. Raufeisen$^c$\\
$^a$Instituto de F\'{\i}sica, Universidade
Federal do Rio Grande do Sul\\ Caixa Postal 15051, CEP 91501-970, Porto
Alegre, RS, Brazil.\\
$^b$Instituto de F\'{\i}sica e Matem\'atica, Universidade Federal de
Pelotas\\
Caixa Postal 354, CEP 96010-090, Pelotas, RS, Brazil\\
$^c$Los Alamos National Laboratory, MS H846, Los Alamos\\ New Mexico
87545, USA.
\end{center}

\begin{abstract}
We investigate the  Drell-Yan transverse momentum
distribution in the framework of the color dipole approach.
Special attention is paid to parton saturation effects at high energies. 
Predictions at LHC energies ($\sqrt{s}=14$ TeV) are given and extrapolated 
down to ISR energies ($\sqrt{s}=62$~GeV).  
Unitarity corrections are implemented through the multiple scattering
Glauber-Mueller approach and are compared  
with predictions of the BGBK saturation model.
\end{abstract}

\section{Introduction}

The high energies available in hadronic collisions at RHIC and to
be reached at LHC will provide new information on parton saturation and 
nuclear phenomena. In these kinematical
domains, massive lepton pairs production in hadronic collisions
(Drell-Yan process \cite{originalDY}) can be used as a clean signal
to investigate the
high parton density regime.

In the color dipole approach, the Drell-Yan (DY) process is viewed in 
the rest frame of the target, where it looks like 
brems\-strah\-lung of a virtual photon decaying into a lepton pair
\cite{bhq}, rather than parton annihilation. 
The DY cross section can then be expressed in terms of the dipole cross
section extracted from small-$x$ Deep Inelastic Scattering (DIS).
Although the mechanism of dilepton production appears to be quite
different in the dipole formulation from what one is used to in the 
parton model, at low $x$,
the equivalence between both approaches has been demonstrated
numerically \cite{Kop1} and analytically \cite{bgdmr}. 
The advantage of the dipole formulation 
is that it allows for an easy implementation
of saturation effects, which avoid the divergence of the DY cross section
at low  
transverse momentum ($p_T\to 0$). 
(in contrast to the result in the infinite momentum frame)

In the following we report the main results on Ref. \cite{bgdmr}, where we  
investigated the DY dilepton
transverse momentum distribution with regards to 
unitarity (parton saturation) aspects, 
which play an important role at high  energies.
These effects are included into
the dipole cross section through
the multiple scattering
Glauber-Mueller approach \cite{AGL}.
Results are compared to predictions from the QCD  improved
saturation model of Ref.~\cite{BGBK}.

\section{The Drell-Yan cross section in the color di\-po\-le approach}

In terms
of color dipole degrees of freedom, the  differential cross
section for radiation of a virtual photon with mass $M$ from a quark 
(or an antiquark)
scattering on a proton  reads \cite{kst1},
\begin{eqnarray}\nonumber\lefteqn{
\frac{d \sigma (qp\to \gamma^* X)}{d\ln\alpha \, d^2p_{T}} }\\
&=&\nonumber\frac{1}{(2\pi)^2}
\int d^2{r_\perp}_1d^2{r_\perp}_2\, 
{\rm e}^{{\rm i}\vec p_T\cdot({\vec r_{\perp1}}-{\vec r_{\perp2}})}
\Psi^{*T,L}_{\gamma^* q}(\alpha,{\vec r_{\perp1}})
\Psi^{T,L}_{\gamma^* q}(\alpha,{\vec r_{\perp2}})\\
&\times&
\frac{1}{2}
\Bigg\{\sigma_{dip}(x_2,\alpha{r_{\perp1}})
+\sigma_{dip}(x_2,\alpha{r_{\perp2}})
-\sigma_{dip}(x_2,\alpha({\vec r_{\perp1}}-{\vec r_{\perp2}}))\Bigg\},
\label{dyptdip}
\end{eqnarray}
where $\alpha$ is the light-cone momentum fraction that the $\gamma^*$
takes from its parent quark and $r_{{\perp}i}$ is related to the
$\gamma^*$-quark transverse separation \cite{kst1}. Explicit expressions for the light-cone
wavefunctions $\Psi^{T,L}_{\gamma^* q}$ can be found {\em e.g.} 
in Ref.~\cite{Kop1}.
The measured cross section
is obtained by embedding 
Eq.~(\ref{dyptdip}) in the hadronic environment, observing that
the projectile quark carries
momentum fraction $x=x_1/\alpha$ 
of the
parent hadron. Correspondingly, $x_1$ is the momentum fraction of
the proton carried by the photon.  
\footnote{We use standard kinematical variables, 
$x_1-x_2=x_F$ and $x_1x_2=(M^2+q_{T}^2)/s$.}

The cross section for a small color dipole scattering off a
nucleon can be
obtained from perturbative QCD. 
However, there is a large uncertainty coming from
non-perturbative aspects (infrared region) of the scattering and
higher orders associated with a perturbative expansion (higher
twists).  A close connection with the DGLAP parton densities can be
obtained in the double logarithmic approximation. In that limit, the
dipole cross section can be written as, 
\vspace{-0.1cm}
\begin{eqnarray}
\sigma_{dip}(x,r_{\perp}) = \frac{\pi^{2} \,
\alpha_s}{3}\,\, r_{\perp}^{2}\,x\,G^{\mathrm{DGLAP}} (x,\tilde{Q}^2)\,,
\label{dglapd}
\end{eqnarray}
where $xG^{\mathrm{DGLAP}} (x,\tilde{Q}^2)$ is the usual DGLAP gluon
distribution at momentum fraction $x$ and virtuality scale
$\tilde{Q}^2=4/r_{\perp}^{2}$ \cite{PRD66}. 
Concerning the non-perturbative contribution, our
procedure is to freeze the dipole cross section in a suitable scale larger than
$r^2_{\mathrm{cut}}$, which corresponds to the initial scale on the
gluon density perturbative evolution $Q^2_0=4/r^2_{\mathrm{cut}}$. 
At high energies, an additional requirement should be obeyed. The growth
of the
partons density (mostly gluons)  has to be tamed, since an
uncontrolled increasing would violate the Froissart-Martin
bound. Then, the black disc limit of the target has to be reached at
quite small Bjorken $x$.  We implement this requirement by using the
multiple scattering Glauber-Mueller approach, which slows down the 
growth of the gluon
distribution in an eikonal way in the impact parameter space
\cite{AGL}.  Therefore, we substitute $xG^{\mathrm{DGLAP}}$ in
Eq. (\ref{dglapd}) by the corrected distribution that also includes unitarity
effects (see Ref.~\cite{PRD66} for further details).
  
The color dipole picture is only valid at small $x_2$, and it
takes into account only sea quarks produced from gluon splitting
in the target, neglecting its valence content. (However, both valence and
sea quarks distributions are parameterized in the 
projectile structure function.)  At lower energies, $x_2$
increases and non-gluonic (valence) contributions to the process are
not negligible. 
In order to extend the dipole approach down to lower energies,
we make use the following educated guess.  The dipole cross section,
Eq.~(\ref{dglapd}), represents the asymptotic gluonic (Pomeron)
contribution to the process. At larger $x$, however,  a
non-asymptotic quark-like content should also be included.
In  Regge language, this contribution corresponds to a Reggeon
instead of a Pomeron exchange in the $t$-channel \cite{PRD66}.  
Hence, we 
add
a Reggeon
contribution 
to the dipole cross section, Eq.~(\ref{dglapd}), 
which is parametrized in a simple way,
$
\sigma^{I\!\!R}_{dip}=\sigma_{0} \,r_{\perp}^2 \,x\,
q_{\mathrm{val}}\,(x,\tilde{Q}^2)\, .
\label{reggnew}
$
With a value of $\sigma_{0}=7$,
we obtain a reasonable description of the DY mass distribution measured by
E772. Similar results were obtained in
Ref.~\cite{PRD66}. The quantity $q_{\mathrm{val}}$ is the valence
quark distribution from the target, evolved through DGLAP.

A quite successful phenomenological realization of the parton
saturation phenomenon is rendered by the QCD improved saturation model
of Ref.~\cite{BGBK}. This model reproduces the DGLAP evolution for
small dipoles and the black disk limit down to small virtualities
(large dipoles) in the eikonal form,
\begin{eqnarray}
\sigma_{dip}(x,r_{\perp})= \sigma_{0}\left\{1-\exp\left(-\frac{\pi^2
r^2_{\perp} \alpha_{s}(\mu^2)xg(x,\mu^2)}{3\sigma_{0}}\right)\right\}\,.
\end{eqnarray}
Here, $g(x,\mu^2)$ is the gluon density of the target, DGLAP evolved
to the scale $\mu^2$, which is assumed to have the form
$\mu^2=\frac{C}{r^2_{\perp}}+\mu_{0}^{2}$. All free parameters have
been taken from the Ref. \cite{BGBK}. In the following we compare the
saturation model with the results coming from the Glauber-Mueller
approach.

\section{Results and Conclusions}

\begin{figure}[!thb]
\vspace*{4.8cm}
\hspace{-0.6cm}
\includegraphics{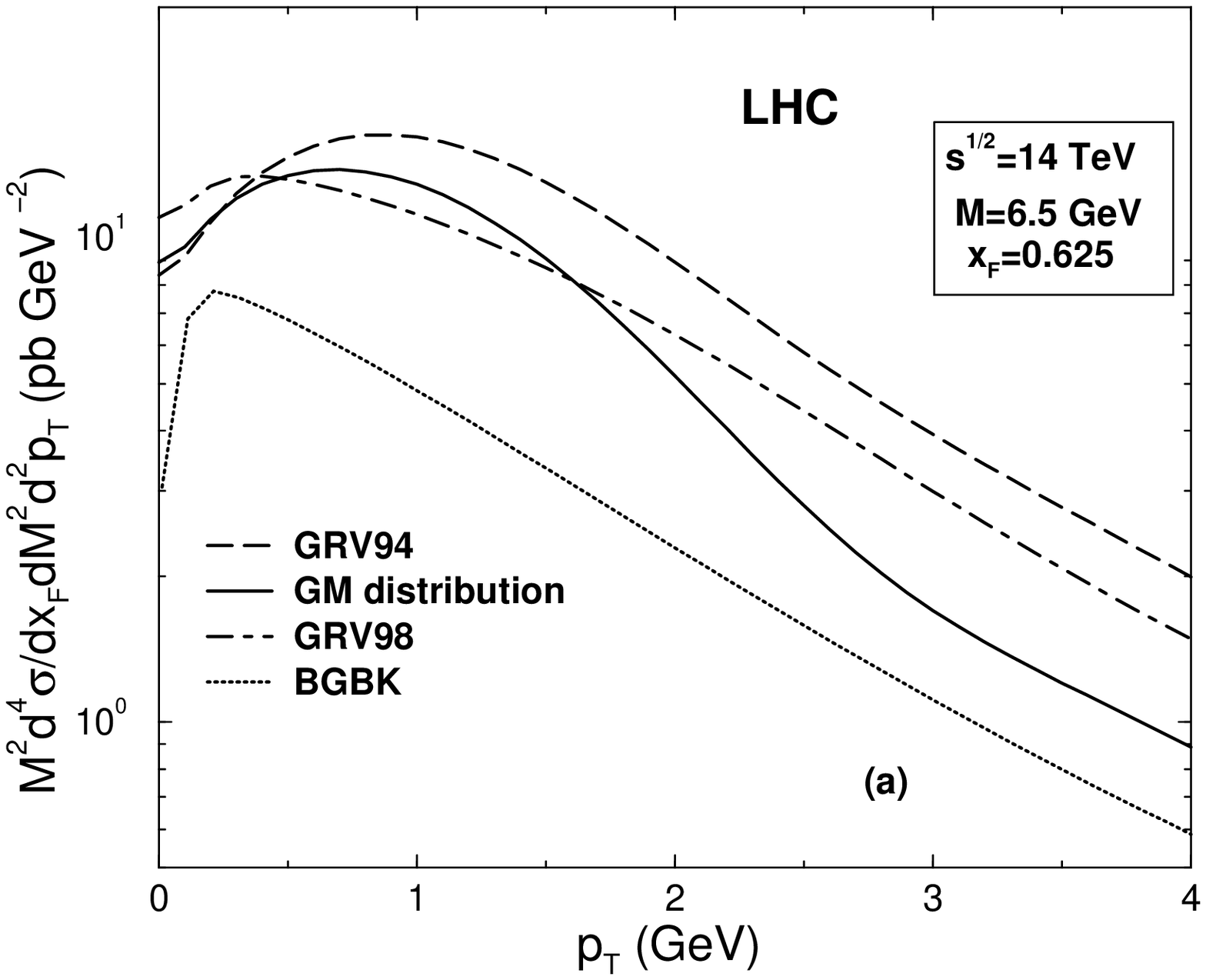} 
\hspace{5.9cm}
\includegraphics{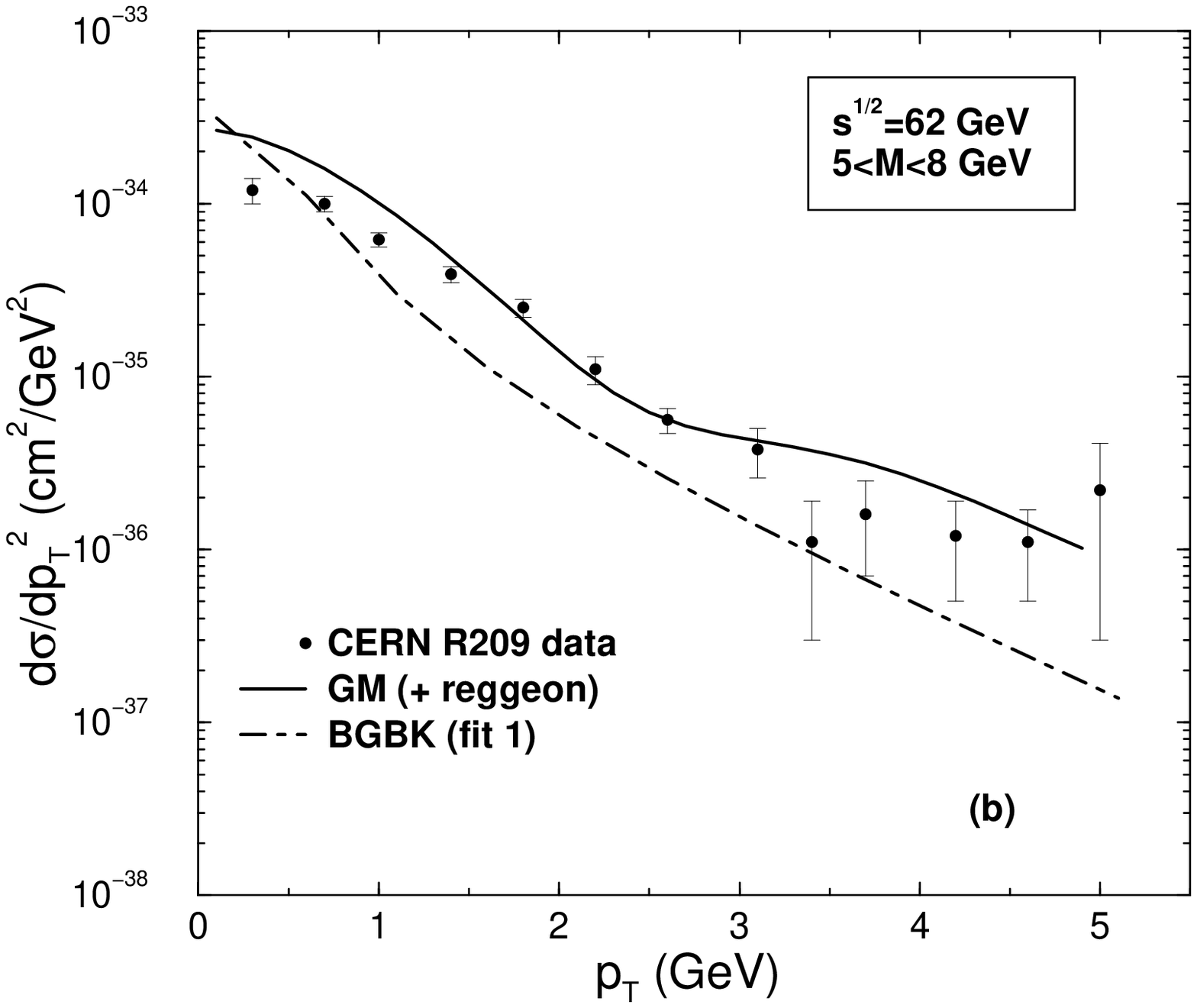}
\vspace*{1cm}
\caption[*]{ The Drell-Yan $p_T$ distribution at (a) LHC
energies (b) and CERNR209  energies $\sqrt{s}=62$ GeV.}
\label{fig1}
\end{figure}

We calculate the DY $p_T$ distribution in proton-proton ($pp$) 
collisions at  LHC energy
($\sqrt{s}=14$ TeV). In addition, we evaluate the
$x_F$-integrated cross section at ISR energy ($\sqrt{s}=62$ GeV)
and  compare it with available DY data (from $pp$ scattering)
in the  mass interval 
$5\leq M\leq 8$ GeV \cite{CERNR209}.
Our results are shown in Fig.~\ref{fig1}.  
The solid line in the Fig.~\ref{fig1}:(a)
denotes the result using the
GM distribution, whereas the dashed line was obtained with the
GRV94 gluon parametrization, {\em i.e.} without saturation. 
The dot-dashed line labels the result using GRV98. The aim of this
comparison is to find out to what extent an updated parametrization could
absorb unitarity effects in the fitting procedure. We conclude that
at LHC energy, those effects could not be absorbed in
a new parametrization.  
As an additional comparison, we present curves from the improved
saturation model BGBK \cite{BGBK} (dotted lines).
In Fig.~\ref{fig1}:(b) the solid curve
denotes the Glauber-Mueller calculation, including the non-asymptotic
valence content (GM + Reggeon) and the dot-dashed line represents 
the BGBK calculations. Both  results are in reasonable agreement with 
shape and overall normalization of the data. 
In order to discriminate among saturation models, one needs 
measurements at high energy accelerators such as RHIC and/or LHC, preferably
in the large rapidity region. 

\end{document}